\documentclass[useAMS,usenatbib]{mn2e}

\usepackage{graphicx,color}

\newcommand{\etal}{{et al.}}
\newcommand{\msun}{\thinspace M_\odot}  
\newcommand{\rsun}{\thinspace R_\odot}

\newcommand{\rhoc}{\rho_{\rm c}}
\newcommand{\nc}{n_{\rm c}}

\newcommand{\cm  }{\,{\rm cm}^{-3} } 

\newcommand{\dfrac}[2]{{\displaystyle \frac{#1}{#2}}  }

\newcommand{\ap }{A_{\phi}}

\newcommand{\omg}{\omega_{\rm c}}
\newcommand{\zsun  }{Z_{\odot} } 
\newcommand{\rs  }{r_{\rm sep} } 
\def\lesssim{\mathrel{\hbox{\rlap{\hbox{\lower4pt\hbox{$\sim$}}}\hbox{$<$}}}}
\def\gtrsim{\mathrel{\hbox{\rlap{\hbox{\lower4pt\hbox{$\sim$}}}\hbox{$>$}}}}

\title[Binary Formation with Different Metallicities]
  {Binary Formation with Different Metallicities: 
Dependence on Initial Conditions}
\author[M. N. ~Machida \etal]
  { Masahiro N. Machida$^{1,2}$\thanks{E-mail: machidam@scphys.kyoto-u.ac.jp (MNM); omukai@th.nao.ac.jp (KO); matsu@i.hosei.ac.jp (TM); inutsuka@tap.scphys.kyoto-u.ac.jp(SI)}, Kazuyuki Omukai$^{2}$, Tomoaki Matsumoto$^{3}$ and Shu-ichiro Inutsuka$^{1}$\\
$^{1}$Department of Physics, Graduate School of Science, Kyoto University, Sakyo-ku, Kyoto 606-8502, Japan\\
$^{2}$National Astronomical Observatory of Japan, Mitaka, Tokyo 181-8588, Japan\\
$^{3}$Faculty of Humanity and Environment, Hosei University, Fujimi, Chiyoda-ku, Tokyo 102-8160, Japan 
}

\pagerange{\pageref{firstpage}--\pageref{lastpage}} \pubyear{2008}

\def\LaTeX{L\kern-.36em\raise.3ex\hbox{a}\kern-.15em
    T\kern-.1667em\lower.7ex\hbox{E}\kern-.125emX}

\begin{document}
\maketitle

\begin{abstract}
The fragmentation process in collapsing clouds with various metallicities is studied using three-dimensional nested-grid hydrodynamics.
Initial clouds are specified by three parameters: cloud metallicity, initial rotation energy and initial cloud shape.
For different combinations of these parameters, we calculate 480 models 
in total and study cloud evolution, fragmentation conditions, 
orbital separation and binary frequency.
For the cloud to fragment during collapse, the initial angular momentum must be higher than a threshold value, 
which decreases with decreasing metallicity.
Although the exact fragmentation conditions depend also on the initial 
cloud shape, this dependence is only modest.
Our results indicate a higher binary frequency in lower-metallicity gas. 
In particular, with the same median rotation parameter as in the 
solar neighbourhood, a majority of stars are born as members of binary/multiple
systems for $< 10^{-4}\zsun$.
With initial mass $<0.1\msun$, 
if fragments are ejected in embryo from the host clouds by multi-body interaction, 
they evolve to substellar-mass objects.
This provides a formation channel for low-mass stars 
in  zero- or low-metallicity environments.
\end{abstract}

\begin{keywords}
binaries: general---cosmology: theory --- galaxies: formation --- hydrodynamics---stars: formation
\end{keywords}

\section{Introduction}
Observations in the solar neighbourhood have shown a high binary frequency 
(\citealt{abt83}; \citealt{duq91}; \citealt{fischer92}; 
see also the review of \citealt{goodwin07}).
Because the binary frequency in star-forming regions is even higher 
than that in the field \citep{mathiu94}, a majority of stars is believed to be born as binaries.
However, the observable star-forming regions are limited in the solar 
neighbourhood and thus at metallicity comparable 
to the solar abundance (i.e., $Z = \zsun$).
We do not know the binary frequency in lower ($Z < \zsun$) 
or zero ($Z=0$) metallicity environments.

Binary stars have important roles in many astrophysical contexts. 
They are considered to be related to a pollution of extremely 
metal-poor stars (Suda et al. 2004; Lucatello et al. 2005; Komiya et al. 2006), progenitors of gamma-ray bursts (Bromm \& Loeb 2006) and Type Ia supernovae (Nomoto, Thielemann, \& Yokoi 1984), the evolution of globular clusters (Sugimoto \& Bettwieser 1983; Makino 1996) and the source of gravitational waves (e.g., Seto 2002).

The formation process of binary (or multiple) systems in collapsing clouds 
with $Z=\zsun$ has been studied by means of three-dimensional 
hydrodynamics 
(see the reviews of \citealt{bodenheimer00} and \citealt{goodwin07}).
They showed that fragmentation occurs after the cloud becomes optically 
thick at $n > 10^{11}\cm$, leading to formation of binary or multiple 
stellar systems.
Whether fragmentation occurs depends sensitively on the initial 
condition, e.g. the initial angular velocity and cloud shapes 
\citep{goodwin07}.
Thus, to determine the fragmentation condition and binary frequency, 
we need to calculate cloud evolution for a number of models.

On the other hand, star formation in primordial clouds has also 
been investigated by cosmological three-dimensional simulations 
\citep{abel02,bromm02,yoshida06}.
Recently, \citet{yoshida08} succeeded in calculating cloud evolution 
up to protostar formation ($n\simeq 10^{21}\cm$) and demonstrated that fragmentation does not occur.
However, cosmological simulations tend to calculate only a single clump 
with the lowest angular momentum, which collapses first. 
This setting prevents the cloud from fragmenting, because binaries form in clouds with a high angular momentum.
Therefore, if we consider not only the clouds collapsing 
first but also those forming at a later time, binaries may form 
within them. 
In fact, \citet{saigo04} and \citet{machida08a} calculated the collapse of primordial clouds with different degrees of initial rotation, and showed 
the possibility of fragmentation and binary formation.
\citet{clark08} demonstrated fragmentation of zero- and low-metallicity 
turbulent clouds by three-dimensional hydrodynamics and showed that the fragments acquire angular momentum from turbulence.

The cloud evolution and fragmentation condition only for primordial clouds were investigated in \citet{machida08a}.
\citet{machida08b} investigated the evolution of clouds with various metallicities and showed that the binary frequency increases as the metallicity decreases.
However, in this study, almost spherical clouds are adopted as the initial states.
As \citet{goodwin07} pointed out, slight differences in the initial conditions may affect cloud evolution and its fragmentation process.
In the present paper, we investigate cloud evolution with various metallicities for a wider range of initial conditions than \citet{machida08b}.
Especially, to investigate the effect of the cloud shape on fragmentation, in some models, we adopt a strongly distorted structure as the initial state.
In total, we have calculated 480 models with combinations of  three parameters: 
the cloud metallicity ($Z$), initial rotation energy ($\beta_0$), 
and initial amplitude of the non-axisymmetric perturbation ($\ap$) 
corresponding to the initial cloud shape.
The large number of models makes it possible to discuss the fragmentation condition and epoch, separation between fragments for clouds with different metallicities.

This paper is organized as follows.
The basics of our model are summarized in \S 2. 
The thermal evolution of clouds with different metallicities is shown in \S3.
Then, we describe the dependence on initial cloud parameters in \S4.
In \S5, we discuss some uncertainties and possible fates of 
the fragments. 
Finally, we summarize our results in \S6.

\section{Model Settings}
\label{sec:model}
We solve the equations of hydrodynamics adopting 
a nested grid method to ensure the Jeans condition of the collapsing cloud. 
Each level of a rectangular grid has the same number of cells of $  128 \times 128 \times 8 $ (for details, see Machida et al. 2005a; 2006a).
The energy equation is not solved: instead, we adopt the barotropic 
relations for pressure as a function of density calculated with 
one-zone models by Omukai et al. (2005). 
Figure~\ref{fig:1} shows these relations for different metallicities as 
a function of the number density.

As an initial condition, we take a spherical cloud with density $1.4$ times 
higher than hydrostatic equilibrium with external pressure 
(i.e. the so-called critical Bonnor--Ebert sphere; 
see \citealt{bonnor56,ebert55}).
The initial central density is taken as $n_{\rm c,0}$ = $1.4\times10^4\cm $.
The critical radius of the Bonnor-Ebert sphere is $R_{\rm c} = 6.45\, 
c_s/[4\pi G \rho_{BE}(0)]^{1/2}$. 
Outside this radius, a uniform gas density 
of $n_{\rm BE}(R_c)=10^3 \cm$ is assumed.
Each cloud rotates rigidly ($\Omega_0$) around the $z$-axis.
The initial temperatures, which are derived from the one-zone model, 
are different in clouds with different metallicities (see  Fig.~\ref{fig:1}).
For example, a cloud with $Z=0$ (primordial composition) has an 
initial temperature of 230\,K, 
while a cloud with $Z=\zsun$ (solar composition) has 7\,K.
As critical Bonnor--Ebert spheres are assumed as the initial state, 
the radii of the initial spheres are different among models with different 
metallicities (or different initial temperatures): 
the radius for models with $Z=0$ is $5.5\times10^5$\,AU, 
while that for models with $Z=\zsun$ is $1.2\times 10^5$\,AU.
We confirmed that initial differences in radius do not greatly 
affect subsequent cloud evolution, in which  
different cloud sizes but the same metallicity are adopted as the initial state.

To induce fragmentation, the non-axisymmetric density perturbation of the $m=2$ mode (i.e. the bar mode) $\delta \rho$ is added to the initial cloud:
\begin{eqnarray}
\delta \rho = \left\{
\begin{array}{ll}
A_{\phi} (r_{\rm c}/r_{\rm p})^2\, {\rm cos}\, 2\phi \; \; \; \mbox{for}\; \;  r_{\rm c} < r_{\rm p}, \\
A_{\phi}\,  {\rm cos}\, 2\phi \; \; \; \mbox{for}\; \;  r_{\rm c} > r_{\rm p},
\label{eq:amp}
\end{array}
\right. 
\end{eqnarray}
where $A_{\phi}$ and $r_{\rm p}$ are the amplitude and effective radius of the density perturbation, respectively, and  $r_{\rm c}$ is the cylindrical radius on the equatorial plane.
We adopt $A_{\phi}$ as $A_{\phi} = 10^{-4}-0.4$, and $r_{\rm p}$  as $0.1\,R_{\rm c}$ and $R_{\rm c}$.
Models with $r_{\rm p}=0.1\,R_{\rm c}$ have a larger non-axisymmetric perturbation than those with $r_{\rm p}=R_{\rm c}$. 
The initial cloud has a strongly distorted shape in models with large $\ap$ and $R_{\rm c}=0.1$.
For convenience, we describe the amplitude of the non-axisymmetric perturbation as $\ap$, in which a capital `L' is added after the value of the amplitude only when $r_{\rm p}= 0.1 R_{\rm c}$ is adopted.
For example, the model with $\ap=0.1$L has parameters of ($A_{\phi}$, $r_{\rm p}$) = (0.1, 0.1$R_{\rm c}$), while that with $\ap=0.1$ has parameters of ($A_{\phi}$, $r_{\rm p}$) = (0.1, $R_{\rm c}$).

The models are characterized by three parameters: the cloud metallicity $Z$, initial rotation energy $\beta_0$ and degree of non-axisymmetric perturbation $\ap$. 
The initial rotation energy means the ratio of rotation to gravitational energy in the initial core.
The values used for these parameters are  $Z$ = $0-1\zsun$, $\beta_0$  = $10^{-6} - 10^{-1}$, and $\ap$ = $10^{-3}-0.3$ and $0.1$L$-$0.4L.
Combining three parameters ($Z$, $\beta_0$, and $\ap$), we calculated 
480 models.

\section{Overview of Thermal Evolution and Its Consequences 
for Prestellar Collapse}
\label{sec:thermal}
\begin{figure}
\begin{center}
\includegraphics[width=150mm]{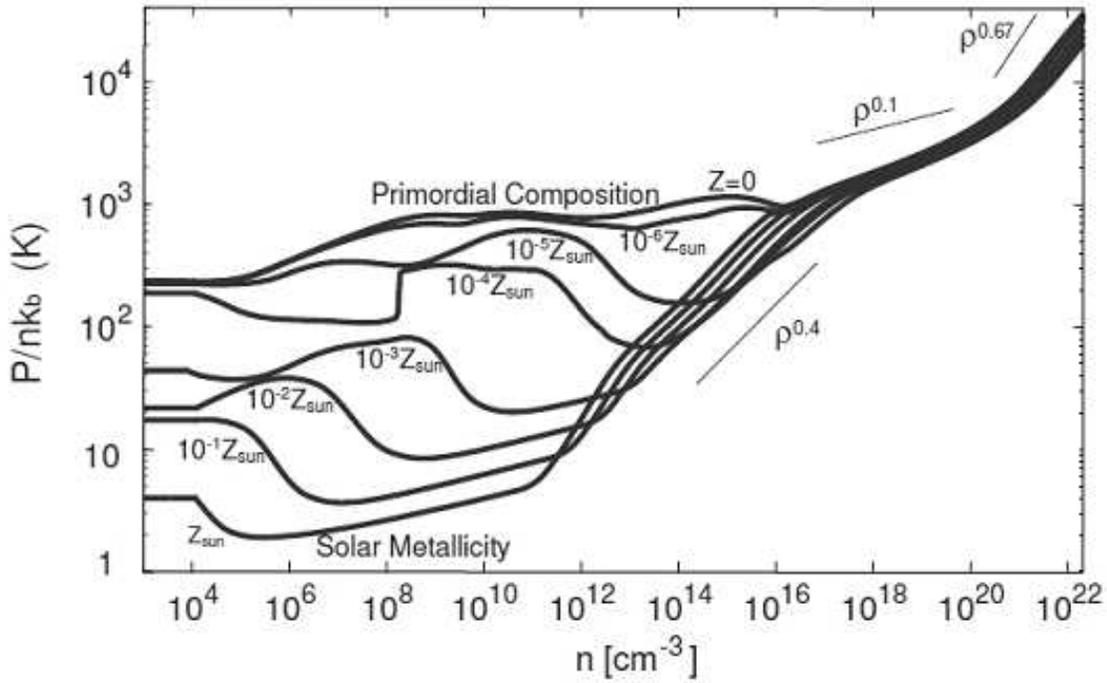}
\caption{
Thermal evolution of collapsing clouds with different metallicities 
($Z = 0$, $10^{-6}$, $10^{-5}$, $10^{-4}$, $10^{-3}$, $10^{-2}$, $10^{-1}$ and $\zsun$) against the number density.
To stress the variation of pressure with density, $P/n k_{\rm b}=T/\mu$ 
is plotted, where $\mu$ is the mean molecular weight.
}
\label{fig:1}
\end{center}
\end{figure}

\begin{figure}
\begin{center}
\includegraphics[width=140mm]{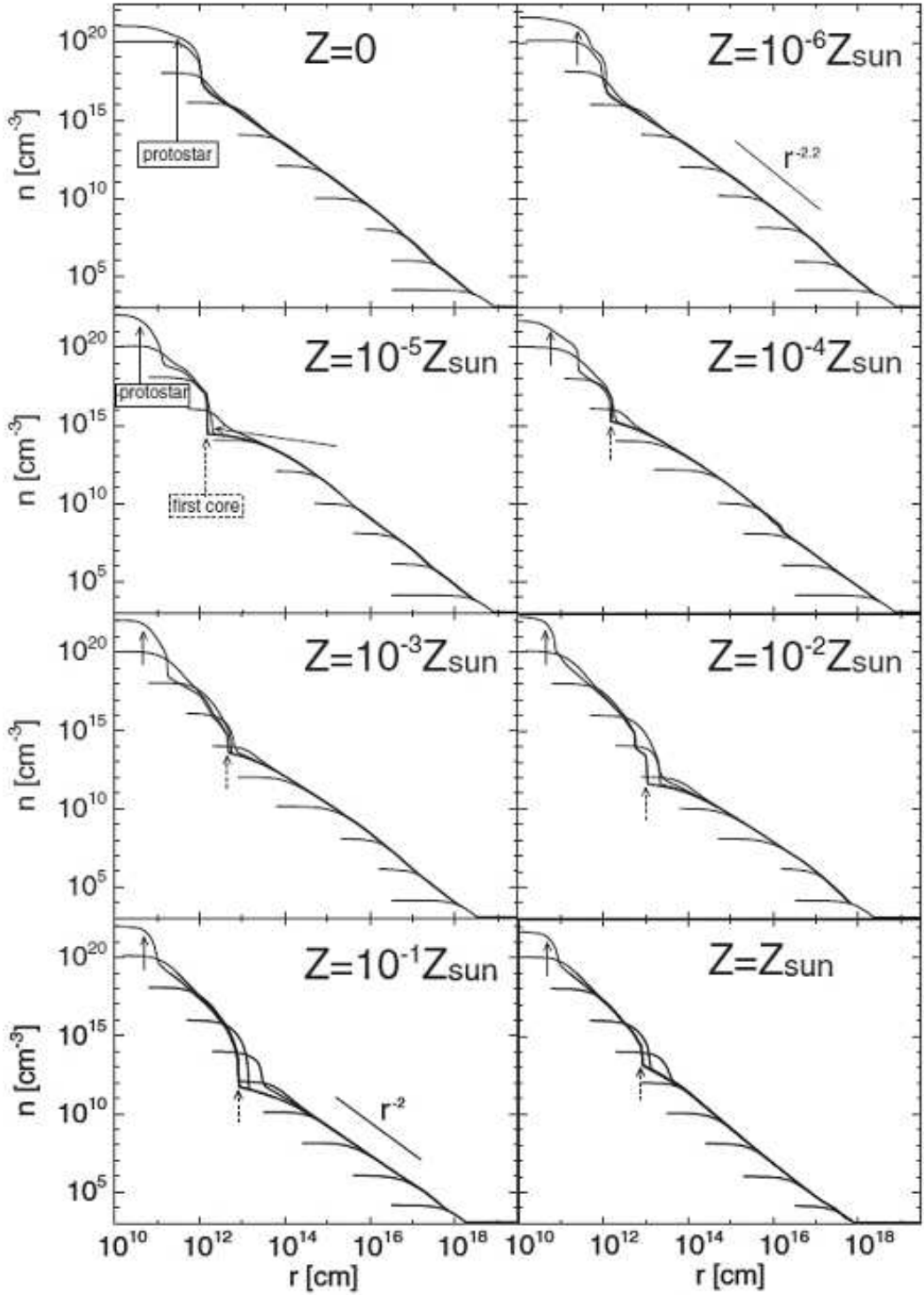}
\caption{
Evolution of radially-averaged density profiles for clouds with 
different metallicity.
Profiles at different times are plotted in the same panel 
for the same cloud.
Formation epochs of the protostar and first core are indicated by 
solid and dotted arrows, respectively.
The relations $r^{-2.2}$ and $r^{-2}$ are also shown in the top right 
and bottom left panels, respectively.
}
\label{fig2}
\end{center}
\end{figure}

\begin{figure}
\begin{center}
\includegraphics[width=140mm]{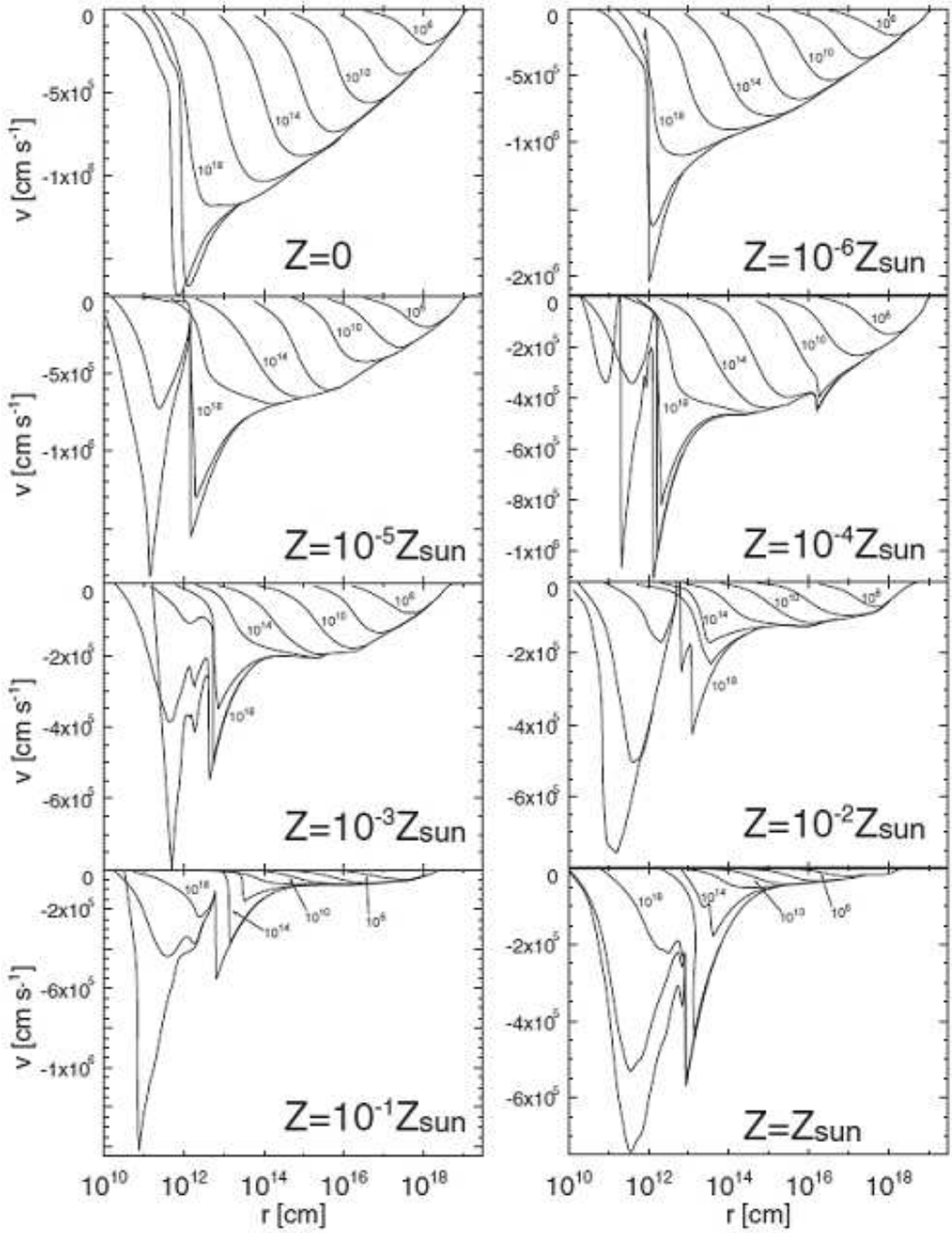}
\caption{
Evolution of radially-averaged velocity profiles for clouds with 
different metallicity.
Numbers in each panel indicate the velocity profiles 
at epochs $\nc=10^6$, $10^{10}$, $10^{14}$ and $10^{18}\cm$.
}
\label{fig3}
\end{center}
\end{figure}

Before discussing fragmentation process, we briefly summarize 
the outline of thermal evolution with different metallicities
(Fig.~\ref{fig:1}), and see its consequence on dynamical 
evolution of clouds with very small rotation energy $\beta_0=10^{-6}$.
For those models, the radial density and velocity profiles
are shown in Figures~\ref{fig2} and \ref{fig3} at 
some different epochs from the initial state ($\simeq 10^4\cm$) 
to after the protostar formation ($\ge 10^{21}\cm$).

The thermal evolution of the clouds with $Z=10^{-6}\zsun$ is 
almost the same as the $Z=0$ case.
For the clouds with $Z=0$ and $Z=10^{-6}\zsun$,
the density and velocity distributions 
are almost the same.
The temperature remains rather high ($\simeq 200$\,K 
at $\simeq 10^4\cm$) and increases gradually with 
the effective ratio of specific heat 
$\gamma \equiv \partial P/\partial \rho \approx 1.1$.
The density gradient in the outer envelope 
is well approximated by $\propto r^{-2.2}$ (Fig.~\ref{fig2}). 
Even though the cloud becomes optically thick to 
the H$_2$ collision-induced absorption at $\simeq 10^{16}\cm$, 
effective cooling by the H$_2$ dissociation prevents the temperature 
from rising adiabatically until 
$\sim 10^{20} \cm$ (Fig.~\ref{fig:1}), where
the dissociation is almost completed.
Subsequently, the growing pressure halts the gravitational collapse at 
the center and the protostar forms, which is surrounded by accretion shock 
($\nc \gtrsim 10^{20}\cm$ at $r\simeq 10^{12}\cm$; Fig.~\ref{fig3}) with
a radius and mass of $\sim 1\rsun$ and $ 10^{-3} \msun $, respectively.

With metal enrichment $Z \ge 10^{-5}\zsun$, a striking difference 
emerges in thermal and dynamical evolution.
The clouds becomes optically thick to the dust in a density around
$10^{11}-10^{15}\cm$, which depends on metallicity in a way that 
those with lower $Z$ becomes adiabatic at a higher density, 
and experience a transient adiabatic phase up to the onset of 
H$_2$ dissociation at $\sim 2000$K 
({\it first adiabatic phase}, hereafter).
Thus, during the first adiabatic phase, 
a transient hydrostatic core, called {\it the first core}, appears 
(indicated by the dotted arrow in Fig.~\ref{fig2}).
The first core has a size of $ \sim 1 $ \,AU, 
and mass of $ \sim 0.0 1\msun $ at its formation.
The central temperature of the first core increases 
with the mass of the first core, which grows by accretion.
At $\sim 2000$K ($\sim 10^{16}\cm$), the H$_2$ begins to dissociate 
and another episode of dynamical collapse begins.
This is called the {\it second collapse phase}, in contrast with the 
{\it first collapse phase} before the first adiabatic phase.
The evolution thereafter coincides with the zero-metallicity case. 
Following the end of dissociation ($\simeq 10^{20} \cm $), 
the temperature increases adiabatically again 
(the {\it second adiabatic phase}) and a second hydrostatic-core 
(the protostar) forms 
with a similar size as that in zero-metallicity case.
Note that two hydrostatic cores (the first core and the protostar) 
appear in a nested manner, i.e., the protostar forms inside the first core.
Correspondingly, two shocks are seen for $Z \ge 10^{-5}\zsun$ 
in Figures~\ref{fig2} and \ref{fig3}.
This existence of two adiabatic phases, 
one at $\nc \sim 10^{11}\cm$ and the other $10^{20}\cm$, during the collapse 
has an important consequence in fragmentation nature of metal-enriched clouds.

In the model with $Z=10^{-4}\zsun$,
three shocks (i.e., three velocity peaks) appear 
with an extra weak shock at 
$\nc \simeq 10^9\cm$ by the rapid H$_2$-formation heating 
at $\nc \simeq 10^{8}\cm$ (see, Fig.~\ref{fig:1}). 
However, because of the very short duration, 
no clear adiabatic core is observed at this epoch. 

\section{Dependence on Initial Cloud Parameters}

In this section, we see the evolution of clouds with 
different values of rotation energy $\beta_0$ and non-axisymmetric 
perturbation $\ap$ (i.e., cloud shape) at the initial states.
For a given metallicity, the number of combination ($\beta_0,\ap$) we 
calculated amounts to 80.
With eight different metallicities $Z=0-\zsun$, 
480 models are calculated in total.
The density distributions on the equatorial plane
at the end of calculation are presented
for different combinations of ($\beta_0,\ap$), which are indicated by 
the ordinate and abscissa, in
Figures~\ref{fig4} - \ref{fig11} for a given metallicity.
The density perturbation is imposed at the scale of 
the critical radius of the BE sphere $R_{\rm c}$ for 
models shown in the first-seventh columns, while 
at $0.1 R_{\rm c}$ for those in the eighth-tenth columns 
(indicated by ``L''; see, \S\ref{sec:model}).
The latter corresponds to models of larger 
non-axisymmetric perturbation (i.e. more distorted).

\subsection{Classification of the Fate of Clouds}
The white-and-black-dotted lines in Figures~\ref{fig4} - \ref{fig11} 
mark the border between models in which the clouds fragment or not.
Each class of models can be classified into two sub-classes, 
which are indicated by the background colours:
the fragmentation models include ``fragmentation'' ({\it red}) 
and ``merger'' ({\it violet}) models, 
while non-fragmentation ones include
``non-fragmentation'' ({\it blue}) and ``stable-core'' ({\it grey}) models.

Fragments survive without merger in ``fragmentation'' models, 
while they merge to form a single core 
before the end of calculation in ``merger'' models.
The protostars form without fragmentation in the ``non-fragmentation'' models, 
while, in the ``stable-core'' models, the first core remains stable 
for $\gg 10\,t_{\rm ff}$, where $t_{\rm ff}$ is the local free-fall 
timescale at the centre.
Due to our CPU-time limitation, we failed to follow further cloud evolution 
for most ``stable-core'' models. 
We expect that the first core collapses eventually to form a protostar 
after angular momentum transfer in a long-term calculation.
In fact, for some ``stable-core'' models, however, we succeeded in following 
the second collapse and protostar formation and confirmed no fragmentation.
Thus, in this paper, we regard the ``stable-core'' models among 
the cases of non-fragmentation when we discuss the fragmentation condition.

\subsection{Fragmentation Frequency for Different Metallicities}
\begin{figure}
\begin{center}
\includegraphics[width=140mm]{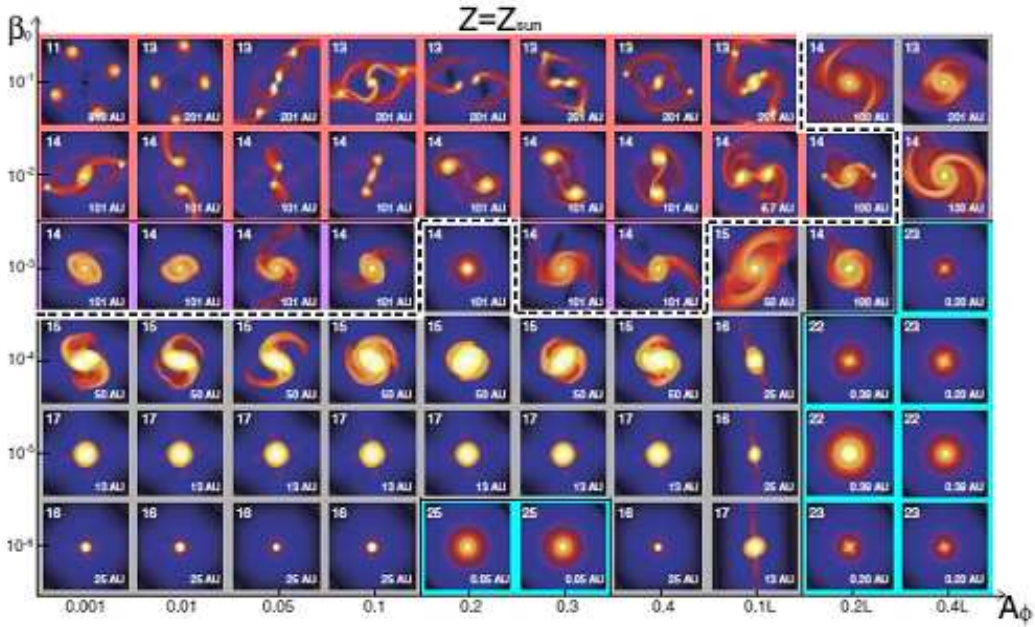}
\caption{
{\small
Final outcomes of cloud collapse for metallicity $Z=\zsun$ 
for different combinations of the initial amplitude of non-axisymmetric 
perturbation $\ap$ and rotation parameter $\beta_0$.
The density distribution ({\it colour-scale}) around the centre of the 
cloud on the equatorial plane is plotted in each panel.
The grid level $l$ and grid scale are shown 
at the upper-left and lower-right corners, respectively, of each panel.
The colours of the panel frame indicate the classifications: 
{\it red}: fragmentation, {\it violet}: merger, {\it blue}: non-fragmentation, and {\it grey}: stable-core models.
The white-and-black dotted line indicates the border between fragmentation (fragmentation and merger models) or not 
(non-fragmentation and stable-core models).
}
}
\label{fig4}
\end{center}
\end{figure}

\begin{figure}
\begin{center}
\includegraphics[width=140mm]{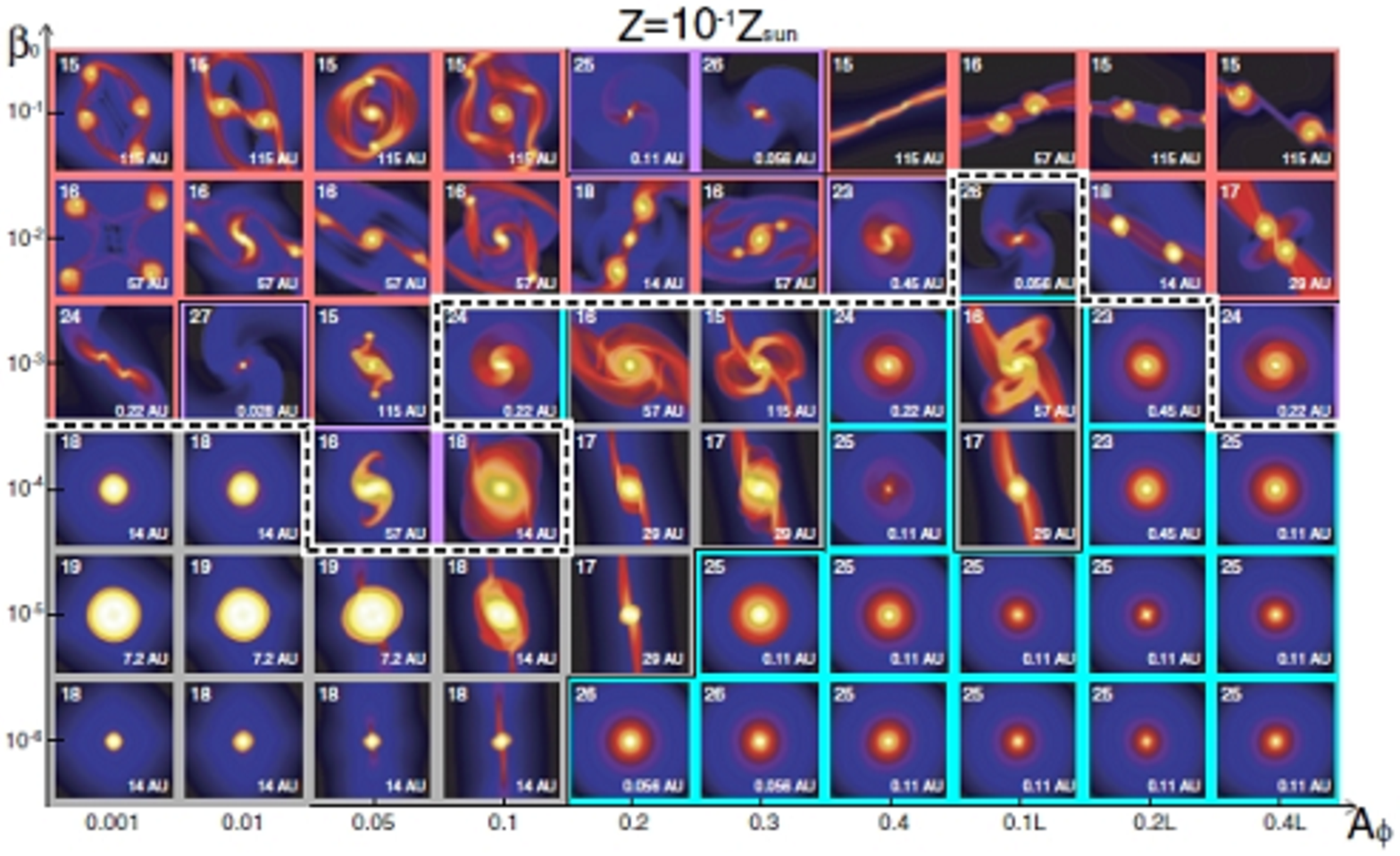}
\caption{
Same as Fig.~\ref{fig4}, but for $Z=10^{-1}\zsun$ with gray-scale.
}
\label{fig5}
\end{center}
\end{figure}

Investigation of Figures~\ref{fig4} - \ref{fig11} indicates:

(i) the first parameter to determine whether the a rotating spherical 
cloud fragments or not is the initial rotation parameter $\beta_{0}$. 
Although the behavior of the boundary between the two cases is very 
complicated, the threshold value of $\beta_{0}$ for fragmentation 
increases with metallicity, being $\beta_0 = 10^{-5}$ for $Z=0$ while 
$\beta_0 = 10^{-3}$ for $Z=\zsun$.
As a result, fragmentation is observed in more models for 
lower metallicity; 
23/60 models for $Z=\zsun$, while 47/60 models for $Z=0$.

(ii) clouds with large non-axisymmetric perturbations develop 
spiral patterns, which transfer angular momenta by gravitational torque 
and effectively reduce the rotation parameters $\beta_0$. 
For example, upward shift of the fragmentation boundary towards 
larger $\ap$ is seen in the $Z=10^{-3}-1 \zsun$ cases.
Similarly, the spirals remove angular momenta from the rotation-supported
first cores, which appears for $0.1$, and $1 \zsun$ 
(Figs.~\ref{fig4} and \ref{fig5}), enabling further collapse 
to the protostar in large $\ap$ cases.
Since the non-axisymmetric perturbation can grow until 
the higher density phase for a cloud with lower metallicity, 
fragmentation tends to occur through the bar configuration. 

\subsection{Spin-up and Fragmentation of Clouds during the Collapse}
\label{sec:frag-cond}
\begin{figure}
\begin{center}
\includegraphics[width=140mm]{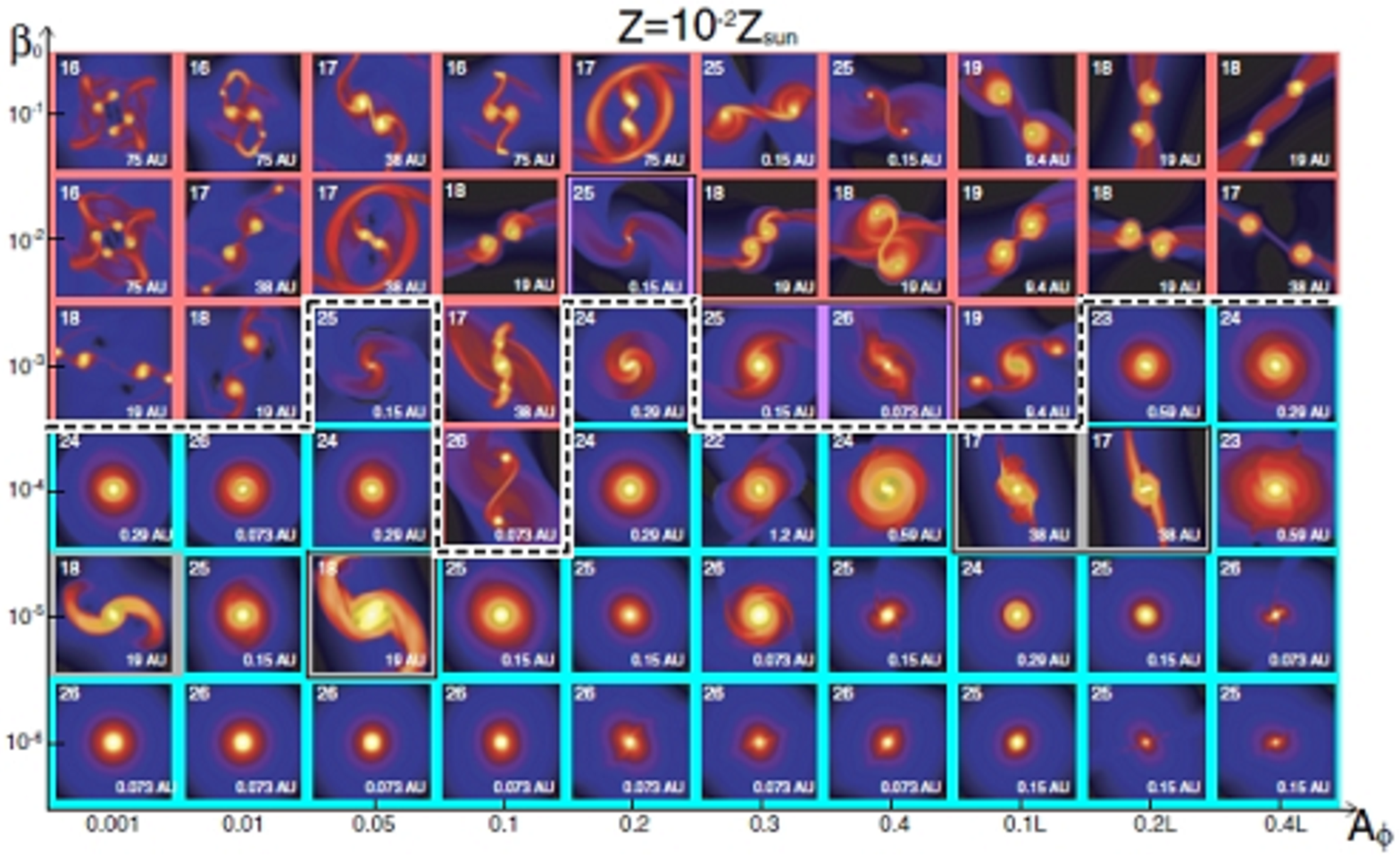}
\caption{
Same as Fig.~\ref{fig4}, but for $Z=10^{-2}\zsun$.
}
\label{fig6}
\end{center}
\end{figure}

\begin{figure}
\begin{center}
\includegraphics[width=140mm]{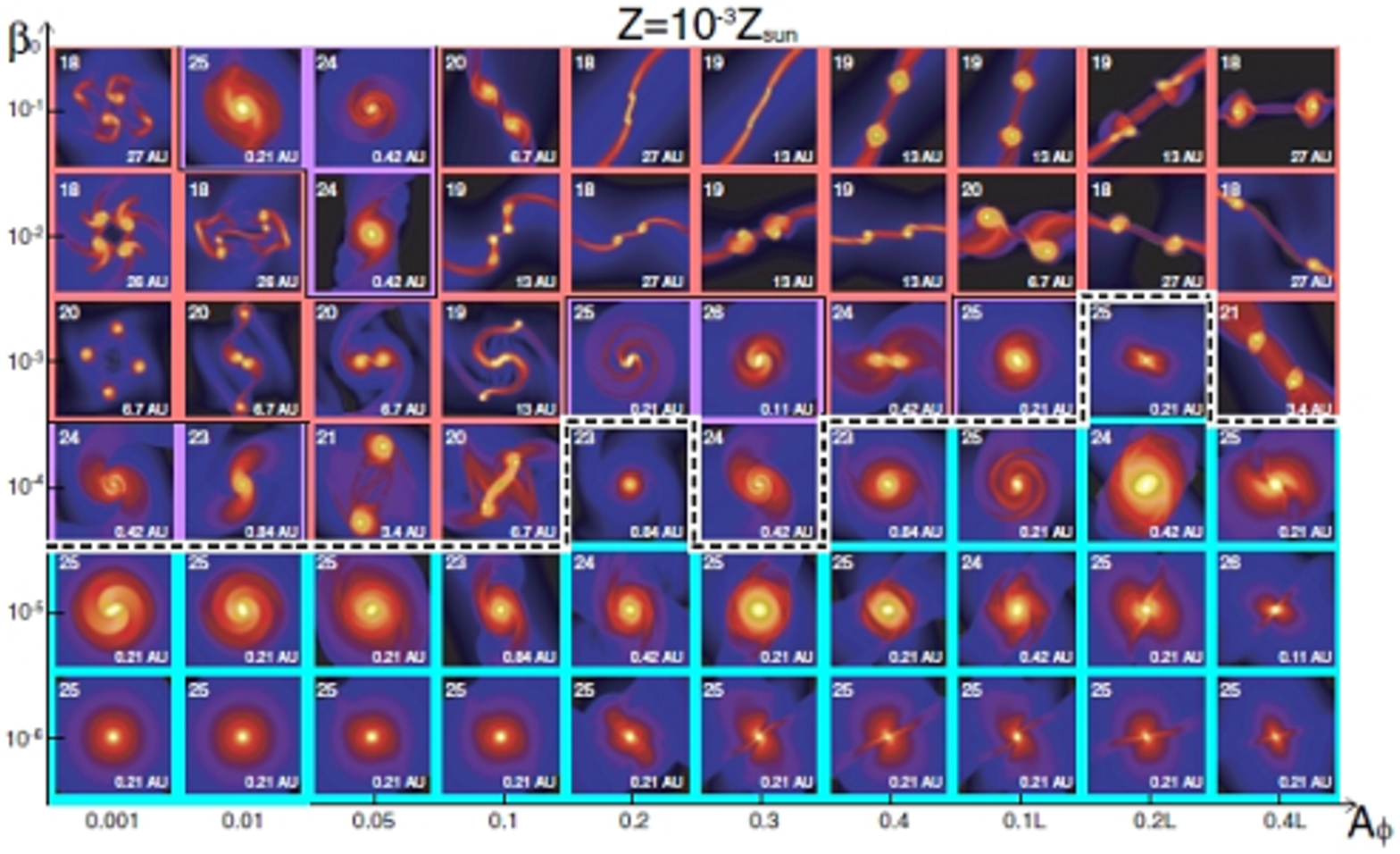}
\caption{
Same as Fig.~\ref{fig5}, but for $Z=10^{-3}\zsun$
}
\label{fig7}
\end{center}
\end{figure}

\begin{figure}
\begin{center}
\includegraphics[width=140mm]{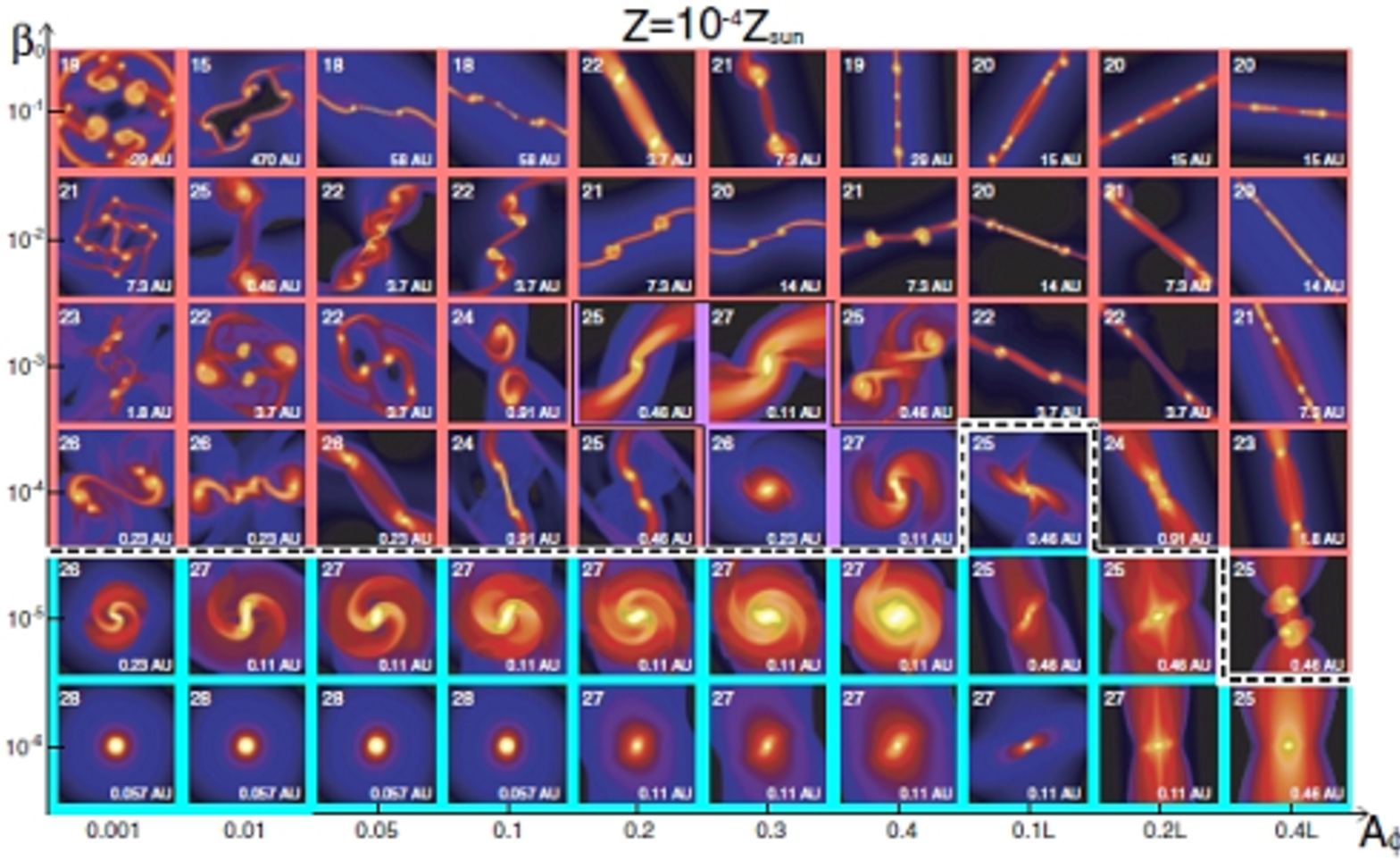}
\caption{
Same as Fig.~\ref{fig4}, but for $Z=10^{-4}\zsun$
}
\label{fig8}
\end{center}
\end{figure}

\begin{figure}
\begin{center}
\includegraphics[width=140mm]{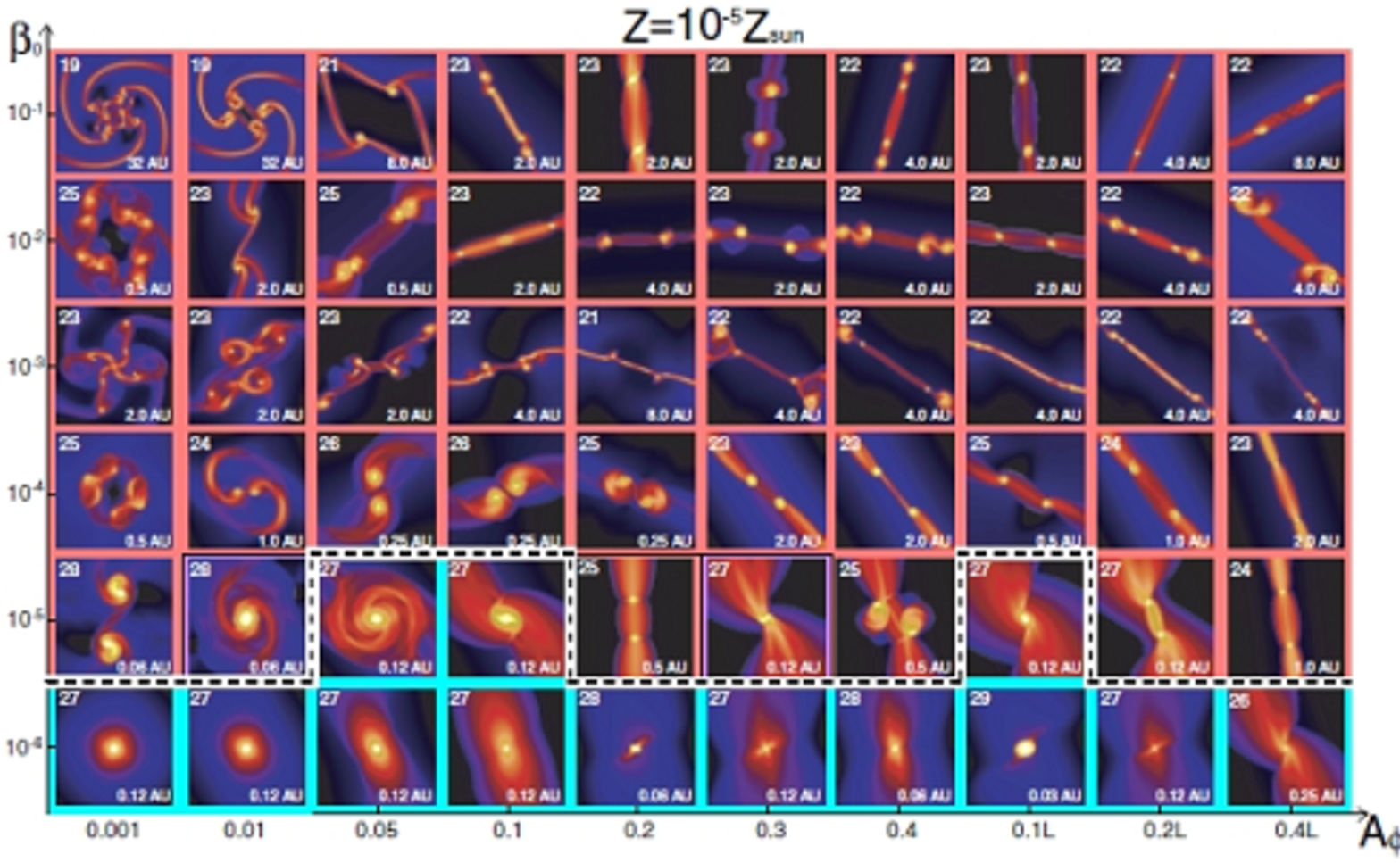}
\caption{
Same as Fig.~\ref{fig5}, but for $Z=10^{-5}\zsun$
}
\label{fig9}
\end{center}
\end{figure}

\begin{figure}
\begin{center}
\includegraphics[width=140mm]{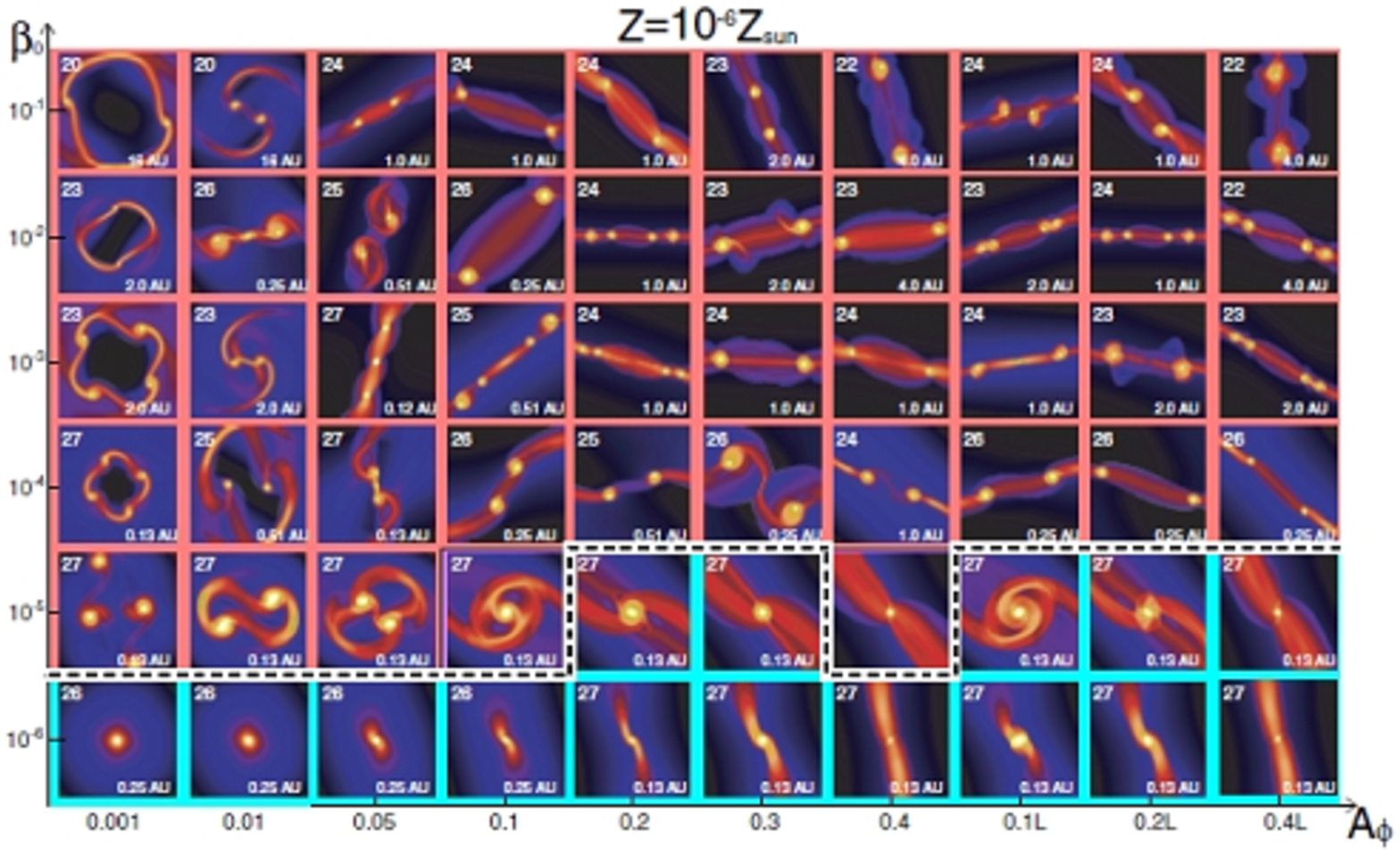}
\caption{
Same as Fig.~\ref{fig4}, but for $Z=10^{-6}\zsun$
}
\label{fig10}
\end{center}
\end{figure}

\begin{figure}
\begin{center}
\includegraphics[width=140mm]{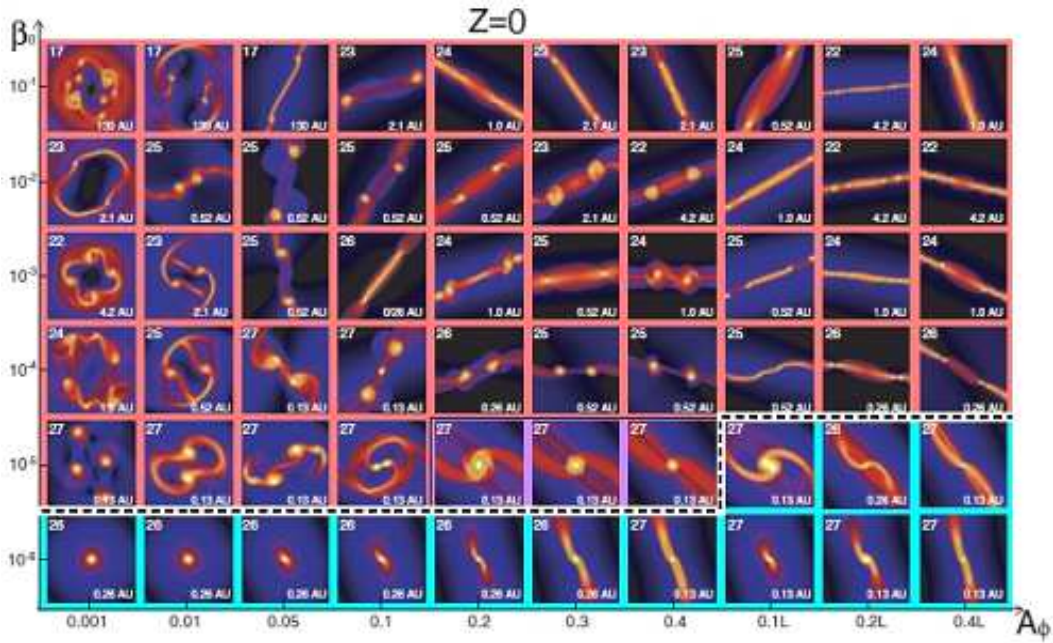}
\caption{
Same as Fig.~\ref{fig5}, but for $Z=0$
}
\label{fig11}
\end{center}
\end{figure}

\begin{figure}
\begin{center}
\includegraphics[width=150mm]{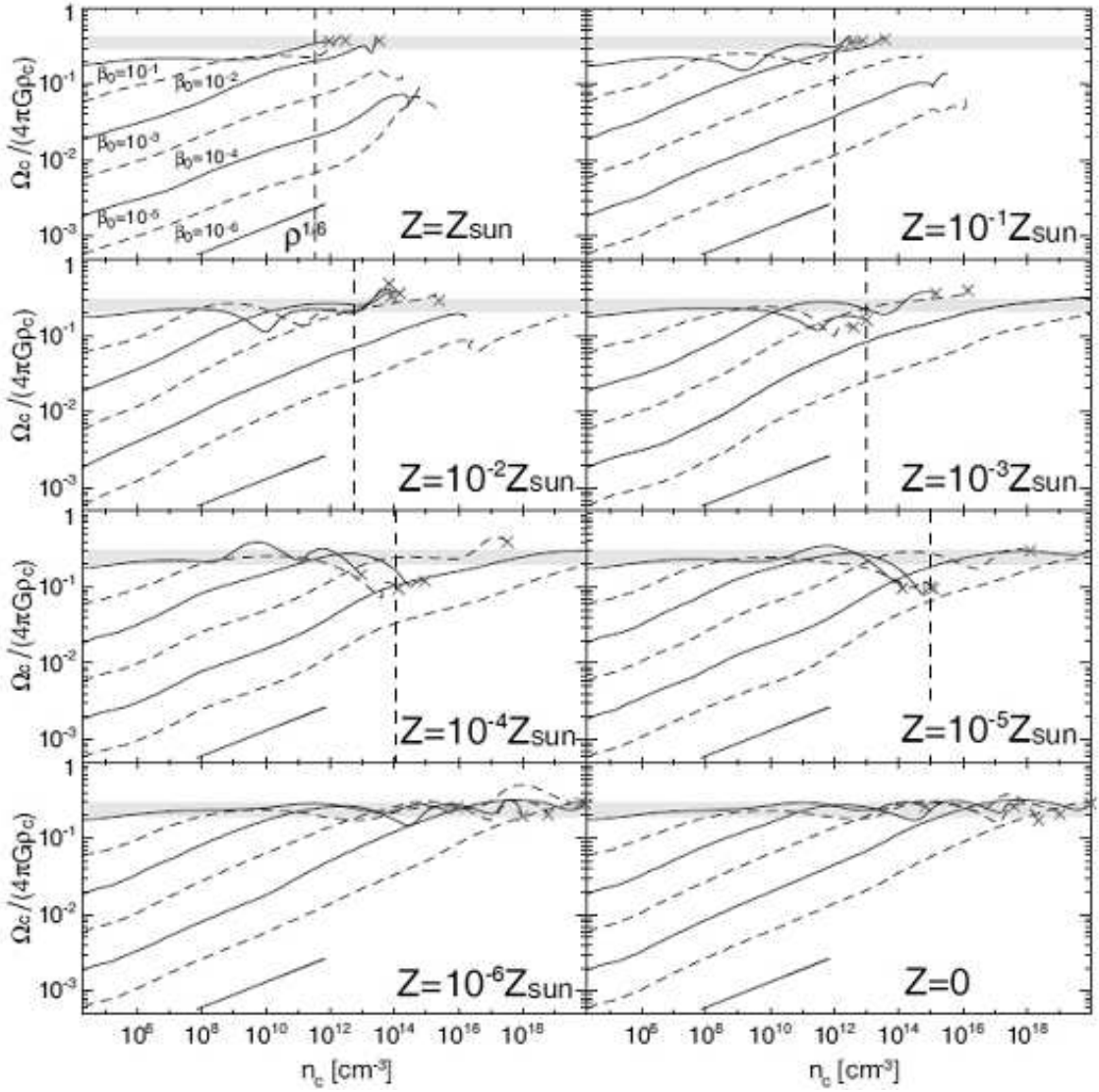}
\caption{
Evolution of the normalized angular velocities at the centre against the number density.
Models with the same $\ap$ (= 0.01) but different $\beta_0$ (=$10^{-1}$--$10^{-6}$) are plotted in the same panels 
for the same metallicities.
The cases with $\beta_0= 10^{-1}, 10^{-3}$ and $10^{-5}$ are indicated 
by solid lines, and those with $\beta_0= 10^{-2}, 10^{-4}$, 
and $10^{-6}$ by dashed lines.
The relation $\propto \rho^{1/6}$, valid for spherical collapse, is also shown by a solid line in each panel.
The vertical broken line indicates the epoch when the gas first becomes adiabatic.
}
\label{fig12}
\end{center}
\end{figure}

As seen in Figures~\ref{fig4} - \ref{fig11}, the 
fragmentation condition depends strongly on the initial rotation rate 
($\beta_0$) but only weakly on the initial cloud shape ($\ap$).
So far, we used the rotation energy of the whole cloud 
for easier comparison with observations.
It is, however, known that a more important index for cloud evolution 
and fragmentation is the angular velocity normalized by the freefall timescale 
(hereafter the normalized angular velocity),
\begin{equation}
\omega_{\rm c} \equiv 
\dfrac{\Omega_{\rm c}}{\sqrt{4\pi\, G\, \rho_{\rm c}}}, 
\end{equation}
where $\Omega_{\rm c}$ and $\rho_{\rm c}$ are the angular velocity and 
density at the centre of the cloud (Matsumoto \& Hanawa 2003, \citealt{machida04}, \citealt{machida05b}).
For a uniform sphere with rigid rotation, 
the normalized angular velocity is related to the rotation energy as
\begin{equation}
\beta_{\rm c} = \dfrac{\Omega_{\rm c}^2 R^3}{3 G M} 
= \dfrac{\Omega_{\rm c}^2}{4\pi G \rhoc}
= \omg^2,
\label{eq:beta}
\end{equation}
where $R$ and $M$ are the radius and mass of the sphere, respectively. 
Along with the collapse, the clouds spin up and $\omega_{\rm c}$ increases
without significant angular momentum transfer.
In the case of the primordial ($Z=0$) clouds,
Machida et al. (2008b) found that 
if the normalized angular momentum reaches the critical value 0.2 - 0.3 (the gray zone in Fig.~\ref{fig12})
and thus a thin disk forms owing to centrifugal force
before the protostar formation, the disk 
subsequently fragments into binary or multiples.
On the other hand, if the cloud becomes 
adiabatic and protostar forms before this condition is met, 
the cloud recovers a spherical shape and does not fragment thereafter.

Here we show that, for higher metallicity,  
appearance of the first core at lower density
put more stringent condition on $\omega_{\rm c}$ 
for fragmentation.
Figure~\ref{fig12} shows evolution of the normalized angular 
velocity $\omega_{\rm c}$ against the central number density for all cases.
Initially, while the centrifugal force is not important, 
the clouds collapse almost spherically and 
$\omega$ (also, $\beta_{\rm c}$) increases in proportion 
to $\propto n_{\rm c}^{1/6}$ 
($\beta_{\rm c} \propto n_{\rm c}^{1/3}$ , respectively).
In clouds with rapid initial rotation, $\omega$ 
reaches the critical value $\omega_{\rm frag} =0.2 - 0.3$ and 
the clouds fragment after some more contraction.
For $Z\le 10^{-6}\zsun$, 
the critical value must be attained 
somewhat before the protostar formation ($\sim 10^{17}{\cm}$) for 
$Z\le 10^{-6}\zsun$, as found 
by Machida et al. (2008b), while for $Z\ge 10^{-5}\zsun$ 
this must be before the first core formation 
(indicated by the vertical lines in Fig.~\ref{fig12}).
Since the first cores form at lower density than the protostars, 
the clouds have a shorter density range to amplify $\omega$ in 
the cases of $Z\ge 10^{-5}\zsun$ than the primordial case.
In addition, first core appears earlier for higher metallicity.
Therefore, an initial higher rotation is required to cause 
fragmentation for clouds with higher metallicity.

Using the relation $\omg \propto n_{\rm c}^{1/6}$ for spherical collapse,
the condition for the normalized angular momentum 
is amplified from the initial value $\omega_{\rm 0}$ at density 
$n_{\rm ini}$
to the critical value $\omega_{\rm frag}$ before the density $n_{\rm adi}$ 
where the gas becomes 
adiabatic is 
\begin{equation}
\omega_{\rm 0} > \omega_{\rm frag} 
\left( 
\dfrac{n_{\rm adi}}{n_{\rm ini}}
\right)^{-1/6}.
\label{eq:omg-cond}
\end{equation}
This can be translated to the condition on initial rotation parameter
$\beta_{\rm 0}$:
\begin{equation}
\beta_{\rm 0} > \beta_{\rm frag} \left( 
\dfrac{n_{\rm adi}}{n_{\rm ini}}
\right)^{-1/3},
\label{eq:frag-cond}
\end{equation}
where $\beta_{\rm frag}= \omega_{\rm frag}^{2}=0.04-0.09$.

The adiabatic density $n_{\rm adi}$ and critical rotation 
parameter $\beta_{\rm 0,crit}$ given by 
the right-hand side of equation~(\ref{eq:frag-cond}) are listed 
in Table~\ref{table:table1} for different values of $Z$.
Here, we used a conservative value of $\beta_{\rm frag}= 0.1$ 
is adopted.
For $Z \ge 10^{-5}\zsun$, $n_{\rm adi}$ is the density where the clouds 
become optically thick to dust grains and the first cores form.
For $Z \le 10^{-6}\zsun$, where no first adiabatic phase is present,
after reaching $\beta_{\rm frag}$, 
the clouds need some more density interval before the protostar 
for fragmentation and thus $n_{\rm adi} = 10^{17}\cm$ is adopted 
as the critical density (see Machida et al. 2008a).
These values for $\beta_{\rm 0, crit}$ are in concordance with the 
results presented in Figures~\ref{fig4} - \ref{fig11}. 

\subsection{Fragmentation Epochs and Separations}

\begin{figure}
\begin{center}
\includegraphics[width=150mm]{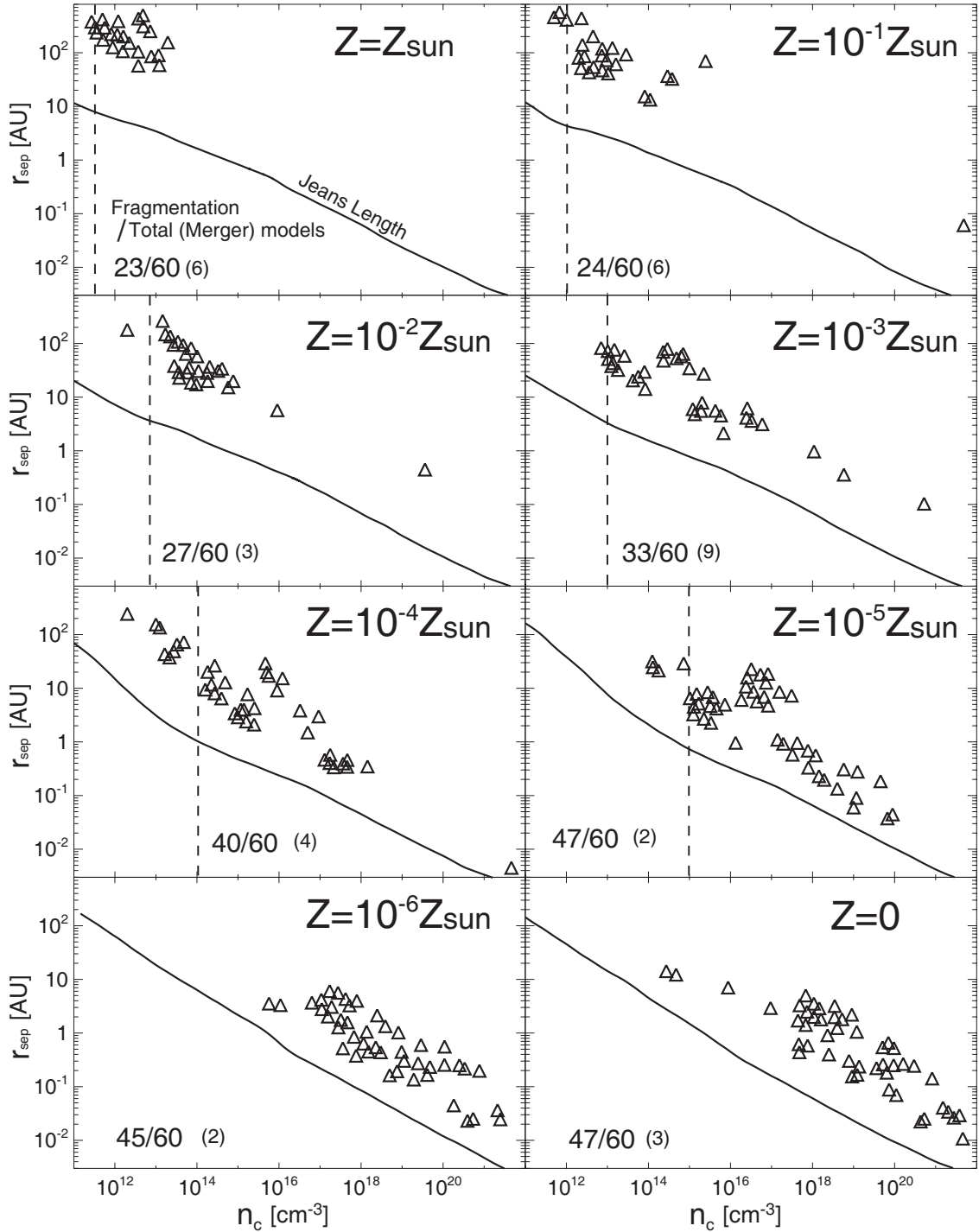}
\caption{
The number density and separation of fragments at the fragmentation epoch.
All fragmentation models are plotted for each metallicity.
The separation corresponds to the furthermost distance between fragments.
The number of fragmentation models is given at the lower left  
in each panel, while the number of merger models is indicated 
by the figure in parentheses.
The Jeans length is also illustrated by solid lines.
The vertical broken line indicates the epoch when the gas first becomes 
adiabatic.
}
\label{fig13}
\end{center}
\end{figure}

Figure~\ref{fig13} shows the separations at the fragmentation epoch 
for all fragmentation models.
In each panel, the solid line indicates the Jeans length, 
while the dashed vertical line indicates the beginning of 
the first adiabatic phase.
Note that the separations are about 10 - 100 times the Jeans length 
since the radial size of the cloud is 10 - 100 times the Jeans scale 
at fragmentation owing to rotation.
With the vertical scale comparable to the Jeans length, 
the height-to-radius ratio is about 1/10 at this epoch.

For $Z=\zsun$, fragmentation events cluster near the upper left corner 
$\nc = 10^{11} - 10^{14} \cm$, i.e., just after first core formation. 
For lower metallicity, this distribution extends toward the lower right
because the first core forms later and thus 
fragmentation occurs at a higher density 
with narrower separation for lower metallicity.
For $Z = 10^{-1} \zsun$ as well as $10^{-2} \zsun$, 
one fragmentation event is observed at very high density 
$\nc \sim 10^{20}\cm$
while the rest locate in $10^{10}\cm < \nc < 10^{14}\cm$, 
which are due to the first-core formation.
In some models of $10^{-4} \zsun$, fragmentation is observed also 
before the first-core formation.
The abrupt temperature rise at $\nc \simeq 10^8\cm$ in the 
$10^{-4} \zsun$ clouds (Figs.~\ref{fig2} and \ref{fig3}) 
temporarily slows the collapse, enabling fragmentation at 
such an early phase.
Without the first-core formation, no clear fragmentation epoch
exists for $Z\le 10^{-6}\zsun$. 
Fragmentation takes place whenever a disk-like configuration 
appears due to rotation and distributes in a wide range 
of $10^{10} \cm < \nc < 10^{21} \cm$ (Machida et al. 2008a)
although some events at $\nc > 10^{20}\cm$ might be caused by 
the protostar formation, as in the models with $Z \ge 10^{-5}\zsun$
at first-core formation.

\begin{figure}
\begin{center}
\includegraphics[width=150mm]{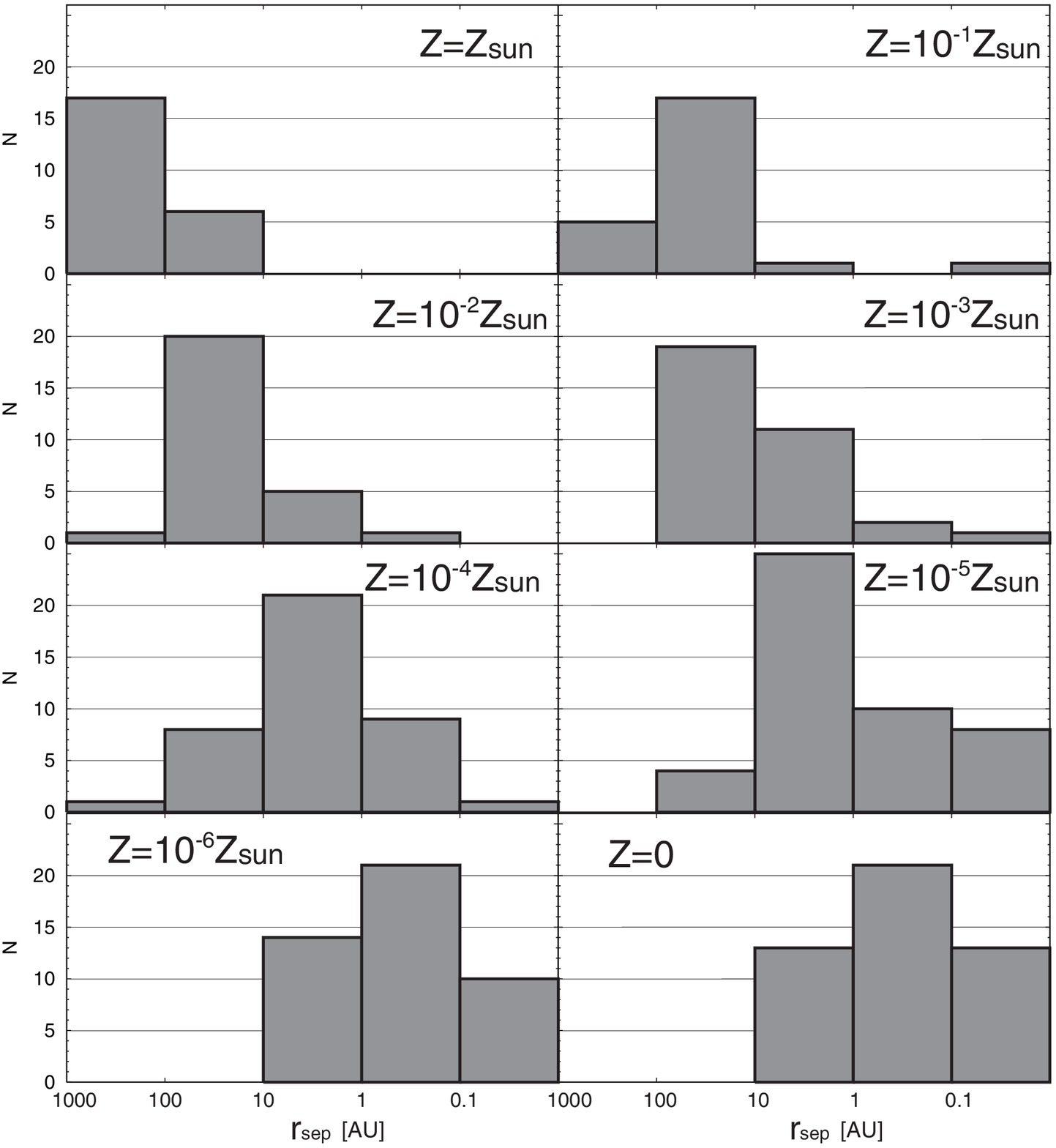}
\caption{
Histogram of the binary separation $\rs$ at fragmentation for all 
fragmentation models. 
The metallicity is indicated in each panel.
}
\label{fig14}
\end{center}
\end{figure}

Figure~\ref{fig14} shows the 
frequency distribution of furthermost separations between fragments 
at fragmentation
This shows that binaries with higher metallicity have a wider separation.
Those with $Z = \zsun$ have separations $10$\,{\rm AU}$ <\rs<1000$\,AU, 
while at $Z = 0$, the range is $\rs < 10$\,AU.
The separations are distributed in a wide range of 
$0.1\,{\rm AU} < \rs < 1000 $\,AU for models with $Z=10^{-2}-10^{-5}\zsun$.

\subsection{Non-axisymmetric Perturbation and Fragmentation Mode}
\label{sec:non-axis}
In a few cases in Figures~\ref{fig4} - \ref{fig11}, the clouds fail
to fragment even if the condition (eq.~\ref{eq:omg-cond}) is fulfilled.
With moderate non-axisymmetry, the cloud can avoid fragmentation even 
if condition  (eq.~\ref{eq:omg-cond}) is fulfilled owing to 
angular momentum is effectively transferred by the spiral arms. 
Interestingly, with even larger $\ap$, the clouds tend to 
fragment through bar configurations instead of ring configurations 
realized for fragmentation in small $\ap$ cases. 
Lower metallicity clouds tend to fragment at higher density.
This allows non-axisymmetric perturbations to grow for a longer time. 
Thus, fragmentation through the bar is more common for lower metallicity.
For example, for $Z=0$, 33 out of total 47 fragmentations 
are through the bar mode.

\section{Discussion}
As shown in \S\ref{sec:frag-cond}, the fragmentation condition is that 
the initial rotation energy $\beta_0$ must be larger than 
the threshold value, which increases with the metallicity.
Thus, with lower metallicities, clouds with smaller rotation energy
can fragment. 
Observations in the solar neighbourhood have shown that molecular clouds 
(with solar metallicity) have rotation energies in the range of 
$10^{-4}<\beta_0<0.07$  with an average value of 
$\beta_0 = 0.02$ \citep{goodman93,caselli02}.
On the other hand, we cannot observe the initial rotation energies 
of clouds with lower or zero metallicity. 
Cosmological simulations have shown that 
host clouds of star formation have rotation energies 
of $\beta \approx 0.1$ \citep{bromm02,yoshida06}, which is slightly larger than those in molecular clouds.
Thus, we expect a higher binary frequency in a lower-metallicity 
environment (or in the early universe), roughly assuming that host clouds 
have almost the same distribution of angular momentum 
(or rotation energy) as observations in the solar neighbourhood. 
If the lower limit of the rotation energy is $\beta_0 \simeq 10^{-4}$, 
as in the solar neighbourhood, fragmentation always occurs in clouds 
with $Z \le  10^{-4} \zsun$, because their fragmentation condition 
is $\beta_0 > 5 \times 10^{-4}$ (see, Table~1).
As a result, the binary frequency increases as the metallicity 
decreases, and most stars are born as binary or multiple systems 
in low-metallicity environments.

In our calculation, fragments have masses of $0.1 - 10^{-3}\msun$ at fragmentation.
This mass range corresponds to that of brown dwarfs or low-mass stars.
The mass of fragments increases in the subsequent accretion phase.
Since the gas supply (or gas accretion) stops at ejection, ejected fragments with mass of $10^{-3}-0.1\msun$ are expected 
to evolve to form a brown dwarf or low-mass star.
Thus, our result indicates that even in extremely low- or zero-metallicity 
environments, substellar-mass objects can form.
On the other hand, fragments remaining in the host cloud are expected to 
increase their mass by accretion and grow to massive stars 
(or massive star binaries).
As a results, we expect that a single massive star 
(or a massive binary system) and multiple low-mass stars 
can be formed in a single host cloud.
When multiple fragments appear like a star cluster, as shown 
in the upper left panels of Figures~\ref{fig4} - \ref{fig11}, low-metallicity stars are expected to have two peaks at $10^{-3}-0.1\msun$ and $100\msun$ in mass distribution.
Note that if fragments stay in the cloud for a while and are ejected after acquiring some mass, a star with mass $M \ge 0.1\msun$ can form.

Subsolar-mass stars can form even in low- (or zero-) metallicity 
environments by the ejection of fragments.
Recently, metal-poor stars with $Z < 10^{-5} \zsun$ are observed 
\citep{christlieb02,frebel05,aoki06}.
Forming in the early universe, they should have a mass of $\lesssim 0.9 \msun$.
Such low-mass metal-poor stars can form by the fragmentation and ejection 
mechanism, as described above.
In addition, our result allows even the existence of zero-metallicity 
stars with sub-stellar mass.
Our calculation showed that the binary frequency for zero-metallicity stars 
is as high as for those with $Z = 10^{-5}$ and $10^{-6}\zsun$.
Thus, we expect observation of zero-metallicity stars in the future.
In addition, as in the cases of ($\ap$, $\beta$) = (0.001, 0.1) 
in Figure~\ref{fig6}, the binary system can be ejected from the host cloud 
when multiple fragments appear.
\citet{suda04} claimed that stars with $Z < 10^{-4} \zsun$ are in 
binaries, because their surface abundance ratios are well explained by mass transfer from an (unobserved) AGB companion. 
The ejection mechanism may explain the origin of such low- or even zero-metallicity binaries.

In this study, we followed the evolution of fragments and protostars only
in their early phase.
To determine the final mass and fate of fragments, 
calculations in the subsequent accretion phase are required.

\section{Summary}
In this research, we studied the binary/multiple formation process
during the collapse of rotating clouds with various metallicities. 
We found the following results:

\begin{enumerate}

\item A transient hydrostatic core (``first core'') appears during the 
prestellar collapse for metallicity $\ge 10^{-5}\zsun$ 
at higher density and with smaller size toward lower metallicity.
For $Z \le 10^{-6}\zsun$, a protostar directly forms without 
any transient core.


\item 
When rotation reaches a fraction of Keplerian value before 
formation of the adiabatic core (i.e., the first core for 
$\ge 10^{-5}\zsun$ or the protostar for $\le 10^{-6}\zsun$),  
the cloud develops disk-like structure and subsequently fragments into 
binary/multiples.
This condition can be translated to the initial rotation 
parameter $\beta_{0}$ exceeding a threshold value $\beta_{\rm 0,crit}$.
Since lower-metallicity clouds have longer density interval to 
spin up, i.e. higher density for the adiabatic-core formation,
they have smaller threshold value.
For example, $\beta_{\rm 0,crit} \sim 10^{-3} $ ($10^{-5}$) 
for $Z = \zsun$ ($Z=0$, respectively).

\item
With modest non-axisymmetric density perturbation, 
angular momentum transfer by gravitational torque
can prevent fragmentation.
However, even larger non-axisymmetry causes fragmentation 
through the bar-mode instability.

\item 
For $Z \ge 10^{-5}\zsun$, majority of fragmentation occur 
just after the first-core formation, while for $Z \le 10^{-6} \zsun$
fragmentation epochs distributes in a wide range of 
$10^{14}\cm < \nc < 10^{22}\cm$.
Clouds with lower metallicity fragments
at a higher density with shorter orbital period and 
their binary separations are narrower. 
Fragmentation proceeds mainly via the bar configuration 
in low ($\lesssim 10^{-3}\zsun$) metallicity clouds, 
while via the ring configuration for higher metallicity.
The difference in fragmentation modes may affect the subsequent 
evolution in the accretion phase.

\item 
With smaller threshold value for the rotation parameter, 
more binaries are expected to form from lower metallicity gas.
With typical rotation rate for nearby star-forming clouds,
most stars with $Z < 10^{-4}\zsun$ are born 
as members of binary/multiple stellar systems.
Fragments have the masses of  $10^{-3}-0.1\msun$ at their formation.
If they are ejected from the host cloud by three-body interaction 
before gaining a large amount of mass, 
substellar-mass objects with extremely low- or even zero-metallicity 
can form.
The observed stars with [Fe/H]$<$-5 might have formed by this mechanism.
Moreover, we expect the possible discovery of zero-metallicity stars
in the Milky-Way halo.
On the other hand, if not ejected, the fragments accrete the ambient gas.
In low-metallicity clouds, because the accretion does not stop until the 
central stars grow to $\gtrsim 100M_{\sun}$ \citep{omukai01,omukai03b,mt08}, 
massive stars or massive binary systems form.
\end{enumerate}

\section*{Acknowledgments}
Numerical computations were carried out on VPP5000 
at National Astronomical Observatory of Japan.
This work was supported by the Grant-in-Aid for the Global COE Program "The Next Generation of Physics, Spun from Universality and Emergence" from the Ministry of Education, Culture, Sports, Science and Technology (MEXT) of Japan, 
and supported in part by the Grants-in-Aid from MEXT 
(18740104, 18740113, 18740117, 19047004 KO, 21740136).

\clearpage
\begin{table}   
\caption{Fragmentation conditions for each metallicity}
\label{table:table1}
\begin{center}
\begin{tabular}{cccccccccccccc}
\hline
Metallicity & $( n_{\rm adi}/10^4 \cm)$ & $\beta_{\rm 0,crit}\, (n_{\rm ini}/10^4 \cm)^{1/3}$ 
\\
\hline
$\zsun$            & $3\times 10^{7}$  & $3.2\times10^{-4}$ \\
$10^{-1} \zsun$    & $10^{8}$          & $2.2\times10^{-4}$ \\
$10^{-2} \zsun$    & $7\times10^{8}$   & $1.1\times10^{-4}$ \\
$10^{-3} \zsun$    & $10^{9}$          & $10^{-4}$ \\
$10^{-4} \zsun$    & $10^{10}$         & $4.6\times10^{-5}$ \\
$10^{-5} \zsun$    & $10^{11}$         & $2.1\times10^{-5}$ \\
$\le10^{-6} \zsun$ & $10^{13}$         & $4.6\times10^{-6}$ \\
\hline
\end{tabular}
\end{center}
\end{table}

\clearpage

\clearpage

\clearpage

\clearpage

\clearpage

\clearpage

\clearpage

\clearpage

\clearpage

\clearpage


\begin{thebibliography}{}{}
\bibitem[\protect\citeauthoryear{Abel \etal}{2002}]{abel02}
Abel, T., Bryan, G. L., \& Norman, M. L. 2002, Science, 295, 93

\bibitem[\protect\citeauthoryear{Abt}{1983}]{abt83}
Abt, H. A. 1983, ARA\&A, 21, 343



\bibitem[\protect\citeauthoryear{Aoki \etal}{2006}]{aoki06} 
Aoki, W., et al., 2006, ApJ, 639, 897 



\bibitem[\protect\citeauthoryear{Bodenheimer \etal}{2000}]{bodenheimer00}
 Bodenheimer P., Burkert A., Klein R. I., \& Boss A. P., 2000, in Mannings V., Boss A. P., Russell S. S., eds, Protostars and Planets IV. Univ. Arizona Press, , p. 675

\bibitem[\protect\citeauthoryear{Bonnor}{1956}]{bonnor56}
 Bonnor, W. B. 1956, MNRAS, 116, 351



\bibitem[\protect\citeauthoryear{Bromm \etal}{2002}]{bromm02}
 Bromm, V., Coppi, P. S., \& Larson, R. B., 2002, ApJ, 564, 23


\bibitem[\protect\citeauthoryear{Bromm \& Loeb}{2006}]{bromm06} 
 Bromm, V., \& Loeb, A. 2006, ApJ, 642, 382


\bibitem[\protect\citeauthoryear{Caselli}{2002}]{caselli02}
 Caselli, P., Benson, P. J., Myers, P. C., \& Tafalla, M. 2002, ApJ, 572, 238 

\bibitem[\protect\citeauthoryear{Christlieb \etal}{2002}]{christlieb02} 
Christlieb, N., et al.\ 2002, Nature, 419, 904 


\bibitem[Clark et al.(2008)]{clark08} 
Clark, P.~C., Glover, S.~C.~O., \& Klessen, R.~S.\ 2008, ApJ, 672, 757 

\bibitem[\protect\citeauthoryear{Duquennoy \& Mayor}{1991}]{duq91}
 Duquennoy, A., \& Mayor, M. 1991, A\&A, 248, 485


\bibitem[\protect\citeauthoryear{Ebert}{1955}]{ebert55}
 Ebert, R. 1955, Z. Astrophys., 37, 222

\bibitem[\protect\citeauthoryear{Fischer \& Marcy}{1992}]{fischer92}
 Fischer, D. A., \& Marcy, G. W. 1992, ApJ, 396, 178

\bibitem[\protect\citeauthoryear{Frebel \etal}{2005}]{frebel05}
 Frebel, A., et al. 2005, Nature, 434, 871

\bibitem[\protect\citeauthoryear{Goodwin \etal}{2007}]{goodwin07}
 Goodwin S. P., Kroupa P., Goodman A., \& Burkert A., 2007, in Reipurth B., Jewitt D., Keil K., eds, Protostars and Planets V. Univ. Arizona Press, , p. 133

\bibitem[\protect\citeauthoryear{Goodman \etal}{1993}]{goodman93}
 Goodman, A. A., Benson, P. J., Fuller, G. A., \& Myers, P. C. 1993, ApJ, 406, 528







\bibitem[\protect\citeauthoryear{Komiya \etal}{2006}]{komiya06}
 Komiya, Y., Suda, T., Minaguchi, H., Shigeyama, T., Aoki, W., \& Fujimoto, M. Y. 2007, ApJ, 658, 367






\bibitem[\protect\citeauthoryear{Machida \etal}{2004}]{machida04}
 Machida, M. N., Tomisaka, K., \& Matsumoto, T. 2004, MNRAS, 348, L1 

\bibitem[\protect\citeauthoryear{Machida \etal}{2005a}]{machida05a}
 Machida, M. N., Matsumoto, T., Tomisaka, K., \& Hanawa, T. 2005a, MNRAS, 362, 369

\bibitem[\protect\citeauthoryear{Machida \etal}{2005b}]{machida05b} 
 Machida, M. N., Matsumoto, T., Hanawa, T., \& Tomisaka, K.  2005b, MNRAS, 362, 382 

\bibitem[\protect\citeauthoryear{Machida \etal}{2006a}]{machida06a} 
 Machida, M. N., Matsumoto, T., Hanawa, T., \& Tomisaka, K.  2006a, ApJ, 645, 1227




\bibitem[\protect\citeauthoryear{Machida et al.}{2008a}]{machida08a} 
 Machida, M.~N., Omukai, K., Matsumoto, T., \& Inutsuka, S.-i.\ 2008a, ApJ, 677, 813 

\bibitem[\protect\citeauthoryear{Machida}{2008b}]{machida08b} 
 Machida, M.~N.\ 2008b, ApJl, 682, L1 


\bibitem[\protect\citeauthoryear{Mathieu}{1994}]{mathiu94} 
 Mathieu, R. D. 1994, ARA\&A, 32, 465



\bibitem[\protect\citeauthoryear{Makino}{1996}]{makino96}
 Makino, J. 1996, ApJ, 471, 796




\bibitem[\protect\citeauthoryear{Matsumoto \& Hanawa}{2003}]{matsu03} 
 Matsumoto T., \& Hanawa T., 2003, ApJ, 595, 913


\bibitem[\protect\citeauthoryear{McKee \& Tan}{2008}]{mt08} 
McKee, C. F., \& Tan, J. C. 2008, ApJ, 681, 771




\bibitem[\protect\citeauthoryear{Nomoto \etal}{1984}]{nomoto84}
Nomoto, K., Thielemann, F.-K., \& Yokoi, K. 1984, ApJ, 286, 644

 

\bibitem[\protect\citeauthoryear{Omukai \etal}{2005}]{omukai05}
 Omukai, K., Tsuribe, T., Schneider, R., \& Ferrara, A. 2005, ApJ, 626, 627

\bibitem[\protect\citeauthoryear{Omukai \& Palla}{2001}]{omukai01} 
 Omukai, K., \& Palla, F.\ 2001, ApJl, 561, L55 

\bibitem[\protect\citeauthoryear{Omukai \& Palla}{2003}]{omukai03b} 
 Omukai, K., \& Palla, F.\ 2003, ApJ, 589, 677







\bibitem[\protect\citeauthoryear{Saigo \etal}{2004}]{saigo04}
 Saigo, K., Matsumoto, T., \& Umemura, M. 2004, ApJ, 615, L65



\bibitem[\protect\citeauthoryear{Sato}{2002}]{sato02}
 Seto, N. 2002, MNRAS, 333, 469

\bibitem[\protect\citeauthoryear{Suda \etal}{2004}]{suda04}
 Suda, T., Aikawa, M., Machida, M. N., Fujimoto, M. Y., \& Iben, I. J. 2004, ApJ, 611, 476 

\bibitem[\protect\citeauthoryear{Sugimoto \& Bettwieser}{1983}]{sugimoto83}
 Sugimoto, D. \& Bettwieser, E. 1983, MNRAS, 204, 19P 














\bibitem[\protect\citeauthoryear{Yoshida \etal}{2006}]{yoshida06}
 Yoshida N., Omukai K., Hernquist L., \& Abel T., 2006, ApJ, 652, 6 

\bibitem[\protect\citeauthoryear{Yoshida \etal}{2008}]{yoshida08} 
Yoshida, N., Omukai, K., \& Hernquist, L.\ 2008, Science, 321, 669 


\end{thebibliography}
\end{document}